\documentclass{article}

\usepackage{arxiv}

\usepackage[utf8]{inputenc} 
\usepackage[T1]{fontenc}    
\usepackage{hyperref}       
\usepackage{url}            
\usepackage{booktabs}       
\usepackage{amsfonts}       
\usepackage{nicefrac}       
\usepackage{microtype}      
\usepackage{graphicx}

\title{Automated Segmentation and Quantification of Choroidal Layers from 3-D Macular OCT Scans}

\author{
 Kyungmoo Lee\\
  Iowa Institute for Biomedical Imaging\\  The University of Iowa\\   \texttt{kyungmoo-lee@uiowa.edu} \\
   \And Alexis K. Warren\\Dept.\ of Ophthalmology and Visual Sciences\\  The University of Iowa\\
   \And Michael D. Abr\`{a}moff\\Dept.\ of Ophthalmology and Visual Sciences\\  The University of Iowa\\
   \And Andreas Wahle\\Iowa Institute for Biomedical Imaging\\  The University of Iowa\\
   \And S. Scott Whitmore\\Dept.\ of Ophthalmology and Visual Sciences\\  The University of Iowa\\
   \And Ian C. Han\\Dept.\ of Ophthalmology and Visual Sciences\\  The University of Iowa\\
   \And John H. Fingert\\Dept.\ of Ophthalmology and Visual Sciences\\  The University of Iowa\\
   \And Todd E. Scheetz\\Dept.\ of Ophthalmology and Visual Sciences\\  The University of Iowa\\
   \And Robert F. Mullins\\Dept.\ of Ophthalmology and Visual Sciences\\  The University of Iowa\\
   \And Milan Sonka\\Iowa Institute for Biomedical Imaging\\  The University of Iowa\\
   \And Elliott H. Sohn\\Dept.\ of Ophthalmology and Visual Sciences\\  The University of Iowa\\   \texttt{elliott-sohn@uiowa.edu}
}

\begin{document}
\maketitle

\begin{abstract}
Background: Changes in choroidal thickness are associated with various ocular diseases, and the choroid can be imaged using spectral-domain optical coherence tomography (SD-OCT) and enhanced depth imaging OCT (EDI-OCT).

New Method: Eighty macular SD-OCT volumes from 80 patients were obtained using the Zeiss Cirrus machine. Eleven additional control subjects had two Cirrus scans done in one visit along with enhanced depth imaging (EDI-OCT) using the Heidelberg Spectralis machine. To automatically segment choroidal layers from the OCT volumes, our graph-theoretic approach was utilized. The segmentation results were compared with reference standards from two independent graders, and the accuracy of automated segmentation was calculated using unsigned/signed border positioning/thickness errors and Dice similarity coefficient (DSC). The repeatability and reproducibility of our choroidal thicknesses were determined by intraclass correlation coefficient (ICC), coefficient of variation (CV), and repeatability coefficient (RC).

Results: The mean unsigned/signed border positioning errors for the choroidal inner and outer surfaces are 3.39 $\pm$ 1.26 $\mu$m (mean $\pm$ standard deviation)/-1.52 $\pm$ 1.63 $\mu$m and 16.09 $\pm$ 6.21 $\mu$m/4.73 $\pm$ 9.53 $\mu$m, respectively. The mean unsigned/signed choroidal thickness errors are 16.54 $\pm$ 6.47 $\mu$m/6.25 $\pm$ 9.91 $\mu$m, and the mean DSC is 0.949 $\pm$ 0.025. The ICC (95\% confidence interval), CV, RC values are 0.991 (0.977–0.997), 2.48\%, 3.15 $\mu$m for the repeatability and 0.991 (0.977–0.997), 2.49\%, 0.53 $\mu$m for the reproducibility studies, respectively.

Comparison with Existing Method(s): The proposed method outperformed our previous method using choroidal vessel segmentation and inter-grader variability.

Conclusions: This automated segmentation method can reliably measure choroidal thickness using different OCT platforms.
\end{abstract}


\section{Introduction}
The choroid is the vascular layer between the retina and the sclera that provides oxygen and nourishment to the outer layers of the retina \cite{Chirco2017,Sohn2014}. Spectral-domain optical coherence tomography (SD-OCT) provides rapid, non-invasive, cross-sectional images of the retina and choroid with high axial resolution and has become a standard clinical and research tool \cite{Mrejen2013,Whitmore2015,Wojtkowski2012}. Visualization of the choroid can be optimized using enhanced depth imaging OCT (EDI-OCT) \cite{Ikuno2011}, which is acquired by placing a zero-delay line to the choroid, or swept-source OCT (SS-OCT) which uses a tunable, longer wavelength laser that enables deeper tissue penetration \cite{Chen2016}. Changes in choroidal thickness are associated with various ocular diseases such as age-related macular degeneration (AMD) \cite{Bakall2013,Manjunath2011,Tozer2013,Whitmore2015}, high myopia \cite{Fujiwara2009}, and central serous chorioretinopathy \cite{Chin2015,Imamura2009,Kim2011,Maruko2010}.

Previous studies have demonstrated the feasibility of manual choroidal layer segmentation of OCT scans \cite{Sacconi2017,Yin2010,Zhang2012}, but manual segmentation is labor intensive and susceptible to effects due to subjectivity of the graders. To overcome these issues, automated methods have been developed using various image processing techniques. Several groups have proposed 2D segmentation methods for choroidal layers of EDI-OCT or SS-OCT scans, but because these algorithms identify choroidal boundaries on each individual B-scan separately, they have the potential to segment discontinuous choroidal layers between adjacent B-scans \cite{Hussain2018,Masood2019,Vupparaboina2015,Zhang2020}. Wang et al. reported an automated 3D segmentation method, which takes into account choroidal layers on adjacent B-scans, but this requires long computational time (43.29 minutes) for each SS-OCT scan \cite{Wang2017}.

Previously, our research group reported an automated 3D choroidal layer segmentation method \cite{Zhang2012} using choroidal vessel segmentation, but it was prone to relatively large segmentation errors since the vessel segmentation is not reliable in low signal choroidal OCT images. Thus, we developed a novel, reliable, memory-efficient, automated 3D choroidal layer segmentation method for macular SD-OCT scans and evaluated it in terms of accuracy, repeatability, and reproducibility.

\section{Materials and Methods}

\subsection{Human Subjects and Data Acquisition}

This study received approval from the institutional review board of the University of Iowa and adhered to the tenets of the Declaration of Helsinki.

To develop this algorithm, subjects were recruited from the glaucoma clinics at the University of Iowa Hospitals and Clinics. Both eyes of all patients had SD-OCT scans but only one eye was randomly chosen from each subject for analysis in this study. Fovea-centered SD-OCT image volumes (200 x 200 x 1024 voxels, 6.0 x 6.0 x 2.0 mm) were acquired from a Cirrus OCT instrument (CirrusTM HD-OCT, Carl Zeiss Meditec, Inc., Dublin, CA) and exported as raw files (.img) using Cirrus Research Browser software. 

To evaluate our choroidal layer segmentation method in terms of accuracy, our results were compared with reference standards created by averaging, in the axial direction, manual tracings ($z$-values in the pixel domain) obtained from two independent graders (AKW, KL). The manual tracings were performed in three B-scan images (horizontal line scans number 50, 100, 150) for each volumetric OCT scan using our custom software which can draw a 2D B-spline using several control points marked by a grader.

Additionally, eleven young subjects without any ocular disease had macular Cirrus SD-OCT scan performed twice in one visit as well as 61 line scan Spectralis (Heidelberg Engineering, Heidelberg, Germany) EDI-OCT scan (768 x 61 x 496 voxels, 9.2 x 7.8 x 1.9 mm) to evaluate our novel segmentation method in terms of repeatability and reproducibility.

\subsection{Automated Segmentation of Choroidal Layers}

\begin{figure}
\centering
\includegraphics[width=1.0\linewidth]{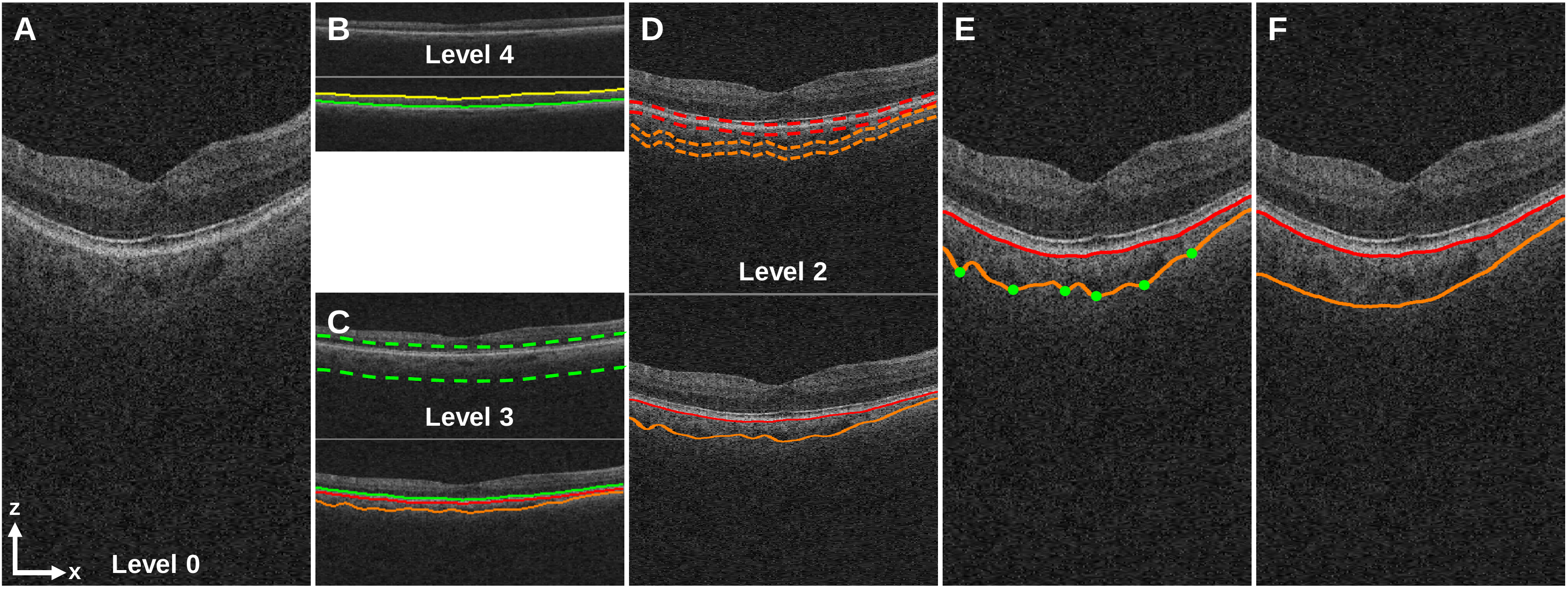}
\caption{Automated multiscale segmentation of the choroidal layer from a 3D macular SD-OCT scan using the LOGIMOS method. (A) Full-resolution central B-scan image (level 0, height: 1024 voxels). (B) Segmentation of the inner limiting membrane (ILM, yellow line) and the boundary of myoid and ellipsoid of inner segments (BMEIS, green line) from the OCT volume (level 4, height: 64 voxels) down-sampled in the $z$-direction by a factor of 16. (C) Segmentation of the BMEIS (green line) and the choroidal layer (red, orange lines) from the OCT volume (level 3, height: 128 voxels) down-sampled by a factor of 8 using the sub-OCT volume (dotted green lines) determined by the BMEIS in level 4 up-scaled by a factor of 2. (D) Segmentation of the choroidal layer from the OCT volume (level 2, height: 256 voxels) down-sampled by a factor of 4 using 2 sub-OCT volumes (dotted red, orange lines) defined by the choroidal layer in level 3 up-scaled by a factor of 2. (E) Segmentation of the choroidal layer from the full-resolution OCT volume (level 0). The green dots show control points for thin plate spline fitting. (F) Image (A) overlaid with the final, smoothed choroidal layer segmentation.}
\label{fig.1}
\end{figure}

The choroidal layer is bounded anteriorly by Bruch’s membrane (BM) and posteriorly by the choroid-sclera interface (CSI). To automatically segment the choroidal layer from a macular SD-OCT image volume, we utilized our layered optimal graph image segmentation for multiple objects and surfaces (LOGISMOS) method \cite{Li2006,Sonka2016,Yin2010} using a multiresolution approach, during which choroidal layer surfaces are sequentially detected at 5 levels of resolution in a coarse-to-fine manner. The analysis starts with level 4 (coarsest, low resolution) and progresses to the finest full resolution level 0. The LOGISMOS method can simultaneously segment multiple surfaces by finding the minimum $s - t$ cut of the weighted graph which is composed of the nodes corresponding to image voxels and the arcs associating graph nodes with costs. Our method starts by segmenting the inner limiting membrane (ILM) and the boundary of myoid and ellipsoid of inner segments (BMEIS) from the macular OCT image volume (level 4) down-sampled in the $z$-direction by a factor of 16 (Fig.~\ref{fig.1}B). To simultaneously detect these surfaces, a double surface 3D graph search approach \cite{Li2006,Sonka2016,Yin2010} was applied using edge-based cost functions of the dark-to-bright transition from top to bottom of the down-sampled OCT image volume. The cost functions are the $z$-directional inverted gradient magnitudes of the OCT voxel intensities, which are smoothed in the $x$-, $z$-directions by a 2D Gaussian filter ($\sigma$ = 1 voxel). Segmentation of the BMEIS, BM, and CSI in level 3 was performed from the sub-OCT image volume region (dotted green lines of Fig.~\ref{fig.1}C), ranging from 30 $\mu$m (2 voxels) above the BMEIS in level 4 up-scaled by a factor of 2 to 400 $\mu$m (26 voxels) below the BMEIS. A triple surface 3D graph search approach \cite{Li2006,Sonka2016,Yin2010} was employed using edge-based cost functions of the dark-to-bright transition for the BMEIS, CSI and that of the bright-to-dark transition for the BM. The $z$-directional first derivatives of the vesselness values obtained by a Hessian matrix analysis \cite{Sato1998} were added to the CSI cost function since the CSI is located below choroid blood vessels. Segmentation of the BM and CSI in level 2 was performed from the sub-OCT image volumes with a height of 86 $\mu$m (11 voxels) (dotted red and orange lines of Fig.~\ref{fig.1}D) based on the BM and CSI segmented in level 3 up-scaled by a factor of 2 using a single surface 3D graph search method \cite{Li2006,Sonka2016,Yin2010}. The BM and CSI in levels 1, 0 were consequently detected using the same manner. Finally, the choroidal layer in the full-resolution OCT image volume was obtained by smoothing the CSI using thin plate spline fitting \cite{Donato2002} (Figs.~\ref{fig.1}E,F). The point (green dot) indicating the thickest location between the BM and CSI surfaces in the patch of 600 $\mu$m (20 voxels) by 600 $\mu$m (20 voxels) was selected, and 100 control points in total were used to create a thin plate spline.

After detecting binary choroidal vessels using a multiscale Hessian matrix analysis followed by vessel probability thresholding, our previous method \cite{Zhang2012} automatically segmented 3D choroidal layers by enveloping the binary choroidal vessels. In addition to an edge-based cost function for the 3D graph search approach, our proposed method utilizes vessel probability values as a vesselness-based cost function to directly segment choroidal layers.

\subsection{Statistical Analysis}
	
To evaluate our automated choroidal layer segmentation method, our results were compared with the reference standards created by averaging in the $z$-direction the manual tracings obtained from two independent graders. The accuracy of our segmentation results was estimated by unsigned/signed border positioning/thickness errors and Dice similarity coefficient (DSC) \cite{Zou2004}. The unsigned/signed border positioning/thickness errors are calculated by measuring the $z$-directional unsigned/signed Euclidean distances between our segmentation results and the reference standards. If our segmented surface is located above the reference standard, the signed border positioning error is negative, otherwise positive. If our choroidal thickness is thinner than the reference standard, the signed thickness error is negative, otherwise positive. The DSC is calculated by measuring the spatial overlap between our segmented choroidal layer (A) and the reference standard (B), which is defined as DSC(A, B) = 2(A $\cap$ B) / (A + B). To validate our choroidal layer segmentation method, the unsigned/signed border positioning/thickness errors and DSCs were compared with the unsigned/signed border positioning/thickness differences and DSCs between the two manual tracings. A paired $t$-test was used in the 95\% confidence interval to compare the two measurements, and a Bland-Altman plot was utilized to describe the agreement between the two measurements \cite{Bland1986}.

The repeatability (between repeated Cirrus SD-OCT scans) and the reproducibility (between Spectralis EDI-OCT and Cirrus SD-OCT scans) of choroidal thicknesses were determined by intraclass correlation coefficient (ICC), coefficient of variation (CV), and repeatability coefficient (RC). The ICC is a statistic representing agreements between two measurements, which is calculated on the basis of a two-way random model for analysis of variance (ANOVA) using statistical software (R version 4.0.0, The R Foundation for Statistical Computing) \cite{Shrout1979}. The CV is defined as a normalized measure of dispersion of two measurements, which is calculated by dividing the standard deviation of two measurements by the mean of the two measurements \cite{Mwanza2010}. The RC is defined as the value of an interval within which 95\% of the differences of two measurements lie, which is measured by multiplying the standard deviation of the differences of the two measurements by 1.96 \cite{Bland1986}.

\section{Results}
	
One hundred Cirrus macular SD-OCT image volumes from 100 patients were randomly obtained from our database system, but 20 volumes were excluded since two independent graders (AKW, KL) could not adequately delineate Bruch’s membrane (BM) or choroid-sclera interface (CSI) because of choroidal neovascularization (CNV), large drusen, and/or fluid. Thus, 80 Cirrus macular SD-OCT image volumes (38 right eyes, 42 left eyes) without significant pathological change in the retina and choroid were used for measurement of the accuracy of our proposed method. The mean age of the cohort was 71.8 $\pm$ 10.9 (mean $\pm$ standard deviation) years, and 34 subjects were male (42.5\%). For measurement of repeatability and reproducibility using our proposed method, two macular Cirrus SD-OCT scans were obtained at the same visit along with Spectralis EDI-OCTs from 11 young subjects (10 female) with a mean age of 31.3 years.

The mean choroidal thicknesses from graders 1, 2, reference standard, our previous \cite{Zhang2012} and proposed methods are 208.75 $\pm$ 54.31 $\mu$m, 206.43 $\pm$ 58.73 $\mu$m, 207.60 $\pm$ 56.00 $\mu$m, 128.18 $\pm$ 49.35 $\mu$m, and 208.85 $\pm$ 54.88 $\mu$m, respectively. There was no significant difference between the two graders ($p = 0.19$) or between the reference standard and our proposed method ($p = 0.26$). The choroidal thickness from our proposed method was significantly thicker than that from our previous method ($p < 0.01$).

\begin{table}
\centering
\caption{Error metrics between our previous vs. proposed choroidal layer segmentation algorithms and the reference standard (RS), and between the manual tracings of two independent graders. The bold font shows a statistically significant improvement of the previous or proposed method ($p < 0.01$).}
\begin{tabular}{ | c | c | c | c | } 
\hline
& Previous Method & Proposed Method & Grader 1 \\ 
& vs. RS & vs. RS & vs. Grader 2 \\ 
\hline
Mean Unsigned Border Positioning Error of BM ($\mu$m)	& 3.85 $\pm$ 3.54 &	3.39 $\pm$ 1.26 &	4.68 $\pm$ 1.61 \\
\hline
Mean Signed Border Positioning Error of BM ($\mu$m)	&	\textbf{0.03 $\pm$ 3.87} &	-1.52 $\pm$ 1.63 &1.08 $\pm$ 2.72 \\
\hline
Mean Unsigned Border Positioning Error of CSI ($\mu$m) & 80.62 $\pm$ 62.50 & \textbf{16.09 $\pm$ 6.21} &	23.73 $\pm$ 10.71 \\
\hline
Mean Signed Border Positioning Error of CSI ($\mu$m) & -79.39 $\pm$ 63.59 &	\textbf{4.73 $\pm$ 9.53} &	-1.24 $\pm$ 15.92 \\
\hline
Mean Unsigned Thickness Error ($\mu$m) & 80.16 $\pm$ 62.73 & \textbf{16.54 $\pm$ 6.47} & 23.90 $\pm$ 10.54 \\
\hline
Mean Signed Thickness Error ($\mu$m) & -79.43 $\pm$ 63.31 & \textbf{6.25 $\pm$ 9.91} &	-2.33 $\pm$ 15.76 \\
\hline
Mean Dice Similarity Coefficient (DSC, unitless) & 0.743 $\pm$ 0.176 & \textbf{0.949 $\pm$ 0.025} & 0.926 $\pm$ 0.035 \\
\hline
\end{tabular}
\label{table.1}
\end{table}

Table~\ref{table.1} shows mean unsigned/signed border positioning/thickness errors and mean DSCs between our previous vs. proposed choroidal layer segmentations and the reference standards, and between the manual tracings of two graders. Our proposed approach showed significantly better performance ($p < 0.01$) than our previous method in all error metrics except the unsigned/signed border positioning errors of BM. There was no significant difference for the unsigned border positioning error of BM ($p = 0.20$), and the previous method showed a significantly smaller signed border positioning error of BM ($p < 0.01$). While the unsigned border positioning errors of BM, CSI, and unsigned thickness error from our proposed approach are significantly smaller than those from the manual tracings of two graders ($p < 0.01$), the signed border positioning errors of BM, CSI, signed thickness error, and DSC from our proposed approach are significantly larger ($p < 0.01$).

\begin{figure}
\centering
\includegraphics[width=1.0\linewidth]{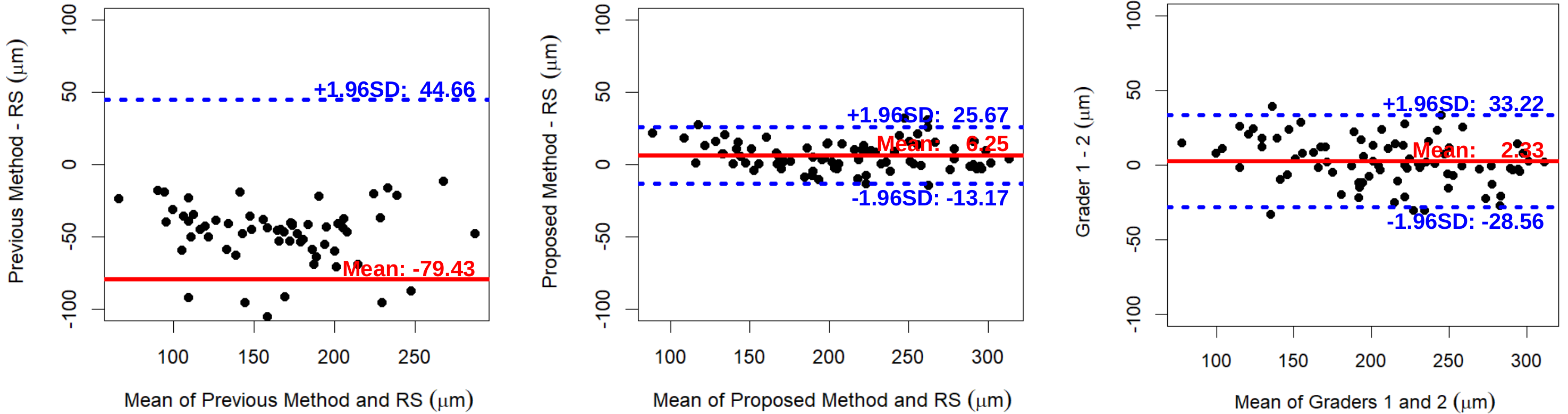}
\caption{Bland-Altman plots of choroidal thicknesses between our previous vs. proposed choroidal layer segmentations and the reference standard (RS) and between the manual tracings of two independent graders for 80 macular SD-OCT scans.}
\label{fig.2}
\end{figure}

As shown in the Bland-Altman plots (Fig.~\ref{fig.2}), the choroidal thicknesses obtained by our previous method are thinner than the reference standards with a larger variation. While the choroidal thickness differences between our proposed method and the reference standard are slightly larger than those between two graders, the proposed method has less variation.

\begin{figure}
\centering
\includegraphics[width=1.0\linewidth]{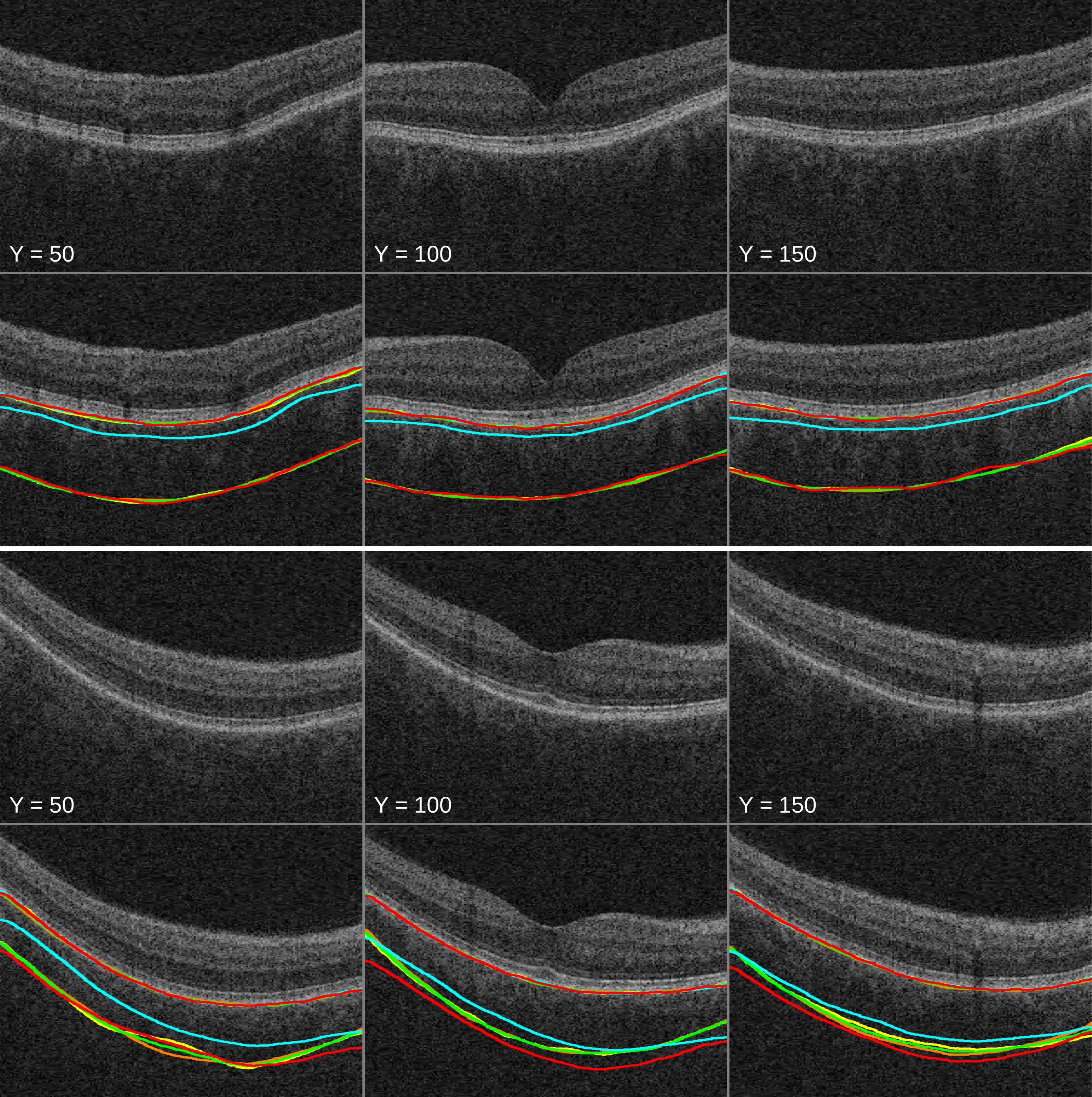}
\caption{Three cropped B-scan images of macular SD-OCT volumes overlaid with our best (top two rows) and worst (bottom two rows) choroidal layer segmentations (yellow line: grader 1, orange line: grader 2, green line: reference standard, cyan line: our previous method, red line: our proposed method).}
\label{fig.3}
\end{figure}

To demonstrate the range of segmentation accuracy across the previous method, proposed method, two graders, and reference standard, Fig.~\ref{fig.3} shows three cropped B-scan images of macular SD-OCT volumes overlaid with our best and worst choroidal layer segmentations having the minimum (4.45 $\mu$m) and maximum (19.30 $\mu$m) sums of the mean unsigned border positioning errors of BM and CSI.

\begin{figure}
\centering
\includegraphics[width=1.0\linewidth]{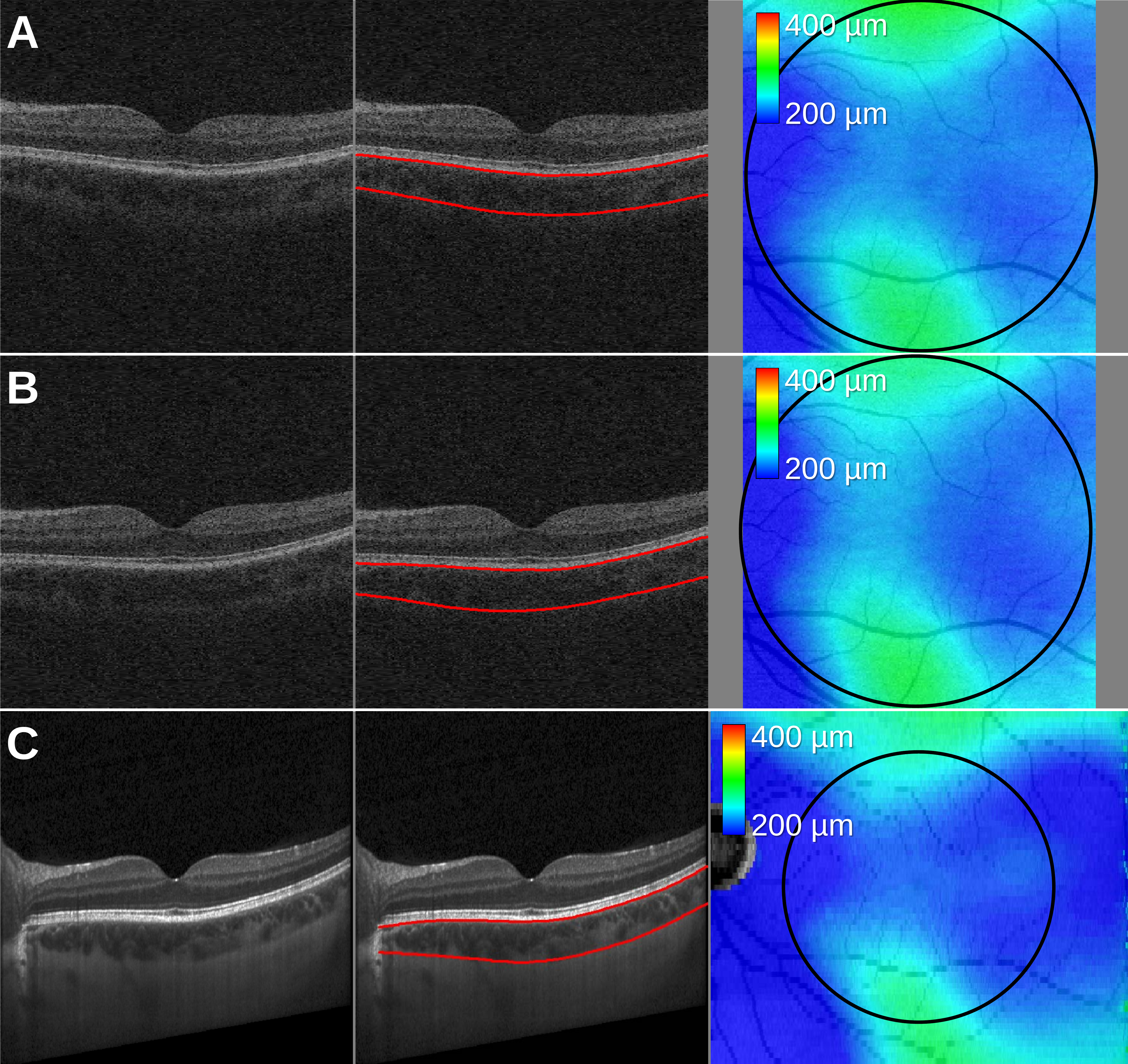}
\caption{Central B-scan images of (A, B) repeated macular Cirrus SD-OCT volumes (6.0 x 6.0 x 2.0 mm) and (C) Spectralis EDI-OCT volume (9.2 x 7.8 x 1.9 mm) of the same eye overlaid with our proposed choroidal layer segmentations (red lines). The black circle on the choroidal thickness map is originated on the fovea, and its radius is 3.0 mm.}
\label{fig.4}
\end{figure}

\begin{table}
\centering
\caption{Intraclass correlation coefficient (ICC), coefficient of variation (CV), and repeatability coefficient (RC) values of our proposed choroidal thicknesses between repeated Cirrus SD-OCT scans for repeatability and between Cirrus SD-OCT and Spectralis EDI-OCT scans for reproducibility.}
\begin{tabular}{ | c | c | c | c | } 
\hline
& SD-OCT 1 & SD-OCT 1 & SD-OCT 2 \\ 
& vs. SD-OCT 2 & vs. EDI-OCT & vs. EDI-OCT \\ 
\hline
Intraclass Correlation Coefficient & 0.991 & 0.991 & 0.987 \\
(ICC, 95\% Confidence Interval) & (0.977–0.997) & (0.977–0.997) & (0.967–0.995) \\
\hline
Coefficient of Variation (CV, \%) & 2.48 & 2.49 & 2.99 \\
\hline
Repeatability Coefficient (RC, $\mu$m) & 3.15 & 0.53 & 3.68 \\
\hline
\end{tabular}
\label{table.2}
\end{table}

\begin{figure}
\centering
\includegraphics[width=1.0\linewidth]{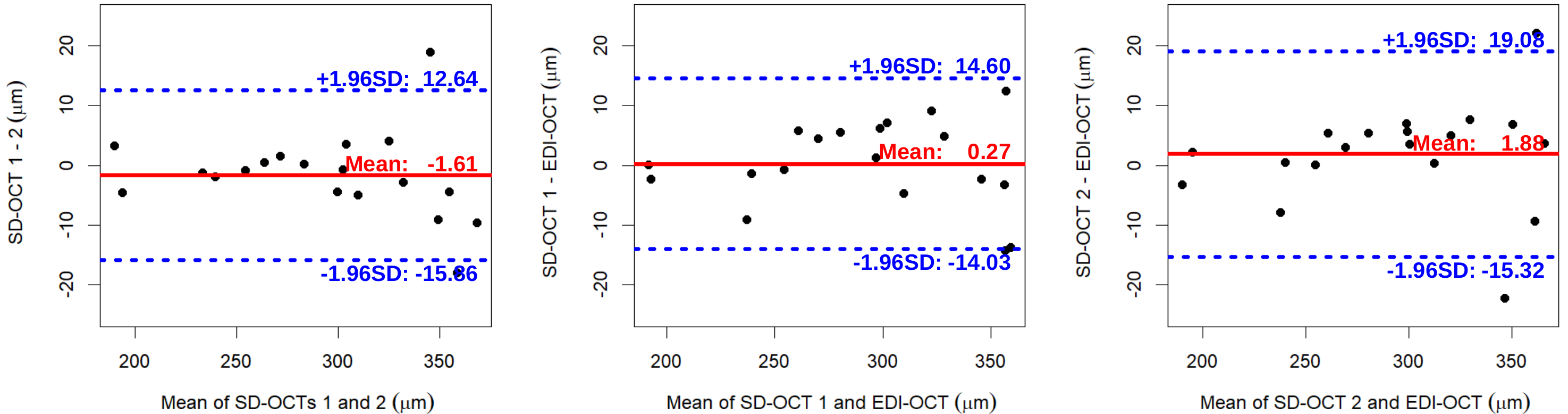}
\caption{Bland-Altman plots of choroidal thicknesses between repeated Cirrus macular SD-OCT scans for repeatability and between Cirrus SD-OCT and Spectralis EDI-OCT scans for reproducibility.}
\label{fig.5}
\end{figure}

The mean choroidal thicknesses of the circular region (radius r $\le$ 3.0 mm) originated on the fovea for the repeated SD-OCT and EDI-OCT scans are 292.73 $\pm$ 53.37 $\mu$m/294.34 $\pm$ 54.89 $\mu$m and 292.46 $\pm$ 53.80 $\mu$m (Fig.~\ref{fig.4}). The mean choroidal thicknesses of the repeated SD-OCT scans are not significantly different ($p = 0.35$), and those of the first/second repeated SD-OCT scans and EDI-OCT scans are not significantly different either ($p = 0.87, 0.36$). Table~\ref{table.2} and Fig.~\ref{fig.5} show the ICC, CV, RC values and Bland-Altman plots of our choroidal thicknesses between the repeated macular SD-OCT scans for the repeatability and between the macular SD-OCT and EDI-OCT scans for the reproducibility.

While our previous method required 6.58 $\pm$ 0.61 minutes of running time and 18.5 GB of computer memory (RAM) per single OCT volume on the PC (OS: Microsoft Windows 10 x64, CPU: Intel® Core™ i7-4790, RAM: 32.0 GB), our proposed method required 1.22 $\pm$ 0.36 minutes and 2.0 GB, notable processing time speedup and decrease of computer memory demands.

\section{Discussion and Conclusions}

Measurement of choroidal thickness, whose changes are associated with various ocular diseases, is clinically important. In this study, we have proposed a novel, reliable, memory-efficient, automated 3D choroidal layer segmentation method for macular SD-OCT scans and evaluated it in terms of accuracy, repeatability, and reproducibility. Our proposed automated choroidal layer segmentation method for 3D macular SD-OCT scans significantly outperformed our previously published method \cite{Zhang2012} using choroid vessel segmentation in terms of mean unsigned/signed border positioning errors of CSI, mean unsigned/signed thickness errors, and mean DSC ($p < 0.01$). There was no significant difference in the BM unsigned border positioning error between our proposed and previous methods ($p = 0.20$), which is not surprising as BM has a clear boundary; thus, even our previous method could detect this well. Compared with BM, the boundary for CSI is less clear, especially on Cirrus scans without dedicated EDI imaging; thus, segmentation of CSI had larger unsigned/signed border positioning errors using our previous method but was substantially improved with the proposed method. As shown in the Bland-Altman plots of Fig.~\ref{fig.2}, the choroidal thicknesses obtained with our proposed method performed better compared with our previous method and also performed well compared with the inter-grader variability of manual segmentations.

Our proposed choroidal layer segmentation method also worked well for macular EDI-OCT scans from a different machine on normal subjects. The ICC, CV, RC values of Table~\ref{table.2} and the Bland-Altman plots of Fig.~\ref{fig.5} showed excellent agreements of our choroidal thicknesses between repeated macular SD-OCT scans for repeatability and between macular SD-OCT and EDI-OCT scans for reproducibility.

Our proposed automated choroidal layer segmentation method was compared with published 2D/3D segmentation methods using EDI-OCT or SS-OCT based on the error metric values provided by their papers. Masood et al. introduced a deep learning-based segmentation method using morphological operations and convolutional neural networks from Spectralis EDI-OCT scans \cite{Masood2019}. The mean unsigned/signed border positioning errors were 5.42 $\pm$ 0.98 $\mu$m/1.68 $\pm$ 3.94 $\mu$m for BM and 11.27 $\pm$ 4.10 $\mu$m/10.92 $\pm$ 5.85 $\mu$m for CSI, and the mean DSC was 0.974 $\pm$ 0.023. While our proposed method showed a better performance than this method in terms of the unsigned/signed border positioning errors of BM and the signed border positioning error of CSI, the deep learning-based method exhibited lower unsigned border positioning error of CSI and the DSC compared with our proposed approach, which is not surprising since the Spectralis EDI-OCT scans depict choroidal layers with less speckle noise and higher contrast than Cirrus SD-OCT scans. Zhang et al. introduced another deep learning-based segmentation method using a biomarker infused global-to-local network from DRI OCT-1 Atlantis SS-OCT scans (Topcon, Tokyo, Japan) \cite{Zhang2020}. The mean unsigned border positioning error of combined BM and CSI was 18.84 $\mu$m, and the mean DSC was 0.908. Our proposed method showed a better performance for both error metrics than this method. Hussain et al. proposed an automated choroidal layer segmentation method using the Dijkstra shortest path algorithm from Spectralis EDI-OCT scans \cite{Hussain2018}. Its mean BM, CSI unsigned border positioning errors, unsigned thickness error, and DSC are 7.29 $\pm$ 2.93 $\mu$m, 30.07 $\pm$ 24.53 $\mu$m, 31.51 $\pm$ 22.97 $\mu$m, and 0.929, respectively. Our proposed method outperformed this method for all error metrics. The Hussain’s method is a 2D segmentation method having a limitation on segmentation of 3D choroidal layers since it does not include the contextual information between adjacent B-scans. Vupparaboina et al. reported another 2D choroidal layer segmentation method from Spectralis EDI-OCT scans by measuring the structural dissimilarity between choroid and sclera by structural similarity (SSIM) index followed by enveloping the thresholded index region \cite{Vupparaboina2015}. Its mean unsigned/signed border positioning errors of CSI and DSC are 19.15 $\pm$ 15.98 $\mu$m, -15.31 $\pm$ 17.97 $\mu$m, and 0.955 $\pm$ 0.017. While this method showed a slightly larger DSC value, our proposed method showed smaller unsigned/signed border positioning errors of CSI. The Vupparaboina’s method used only 5 EDI-OCT scans for validation and required 6-12 minutes per scan of running time, which is 4.9-9.8 times slower than our proposed method. Want et al. presented an automated 3D segmentation method using the level set framework from DRI OCT-1 Atlantis SS-OCT scans \cite{Wang2017}. Its mean unsigned/signed border positioning errors of CSI and DSC are 5.64 $\pm$ 4.60 $\mu$m, 4.13 $\pm$ 4.29 $\mu$m, and 0.90 $\pm$ 0.04. While this method showed smaller unsigned/signed border positioning errors of CSI, our proposed method showed a larger DSC value. The SS-OCT scans also depict choroidal layers with higher-than-Cirrus contrast, similar to the EDI-OCT scans. The drawback of this method is its long computational time (43.29 minutes) compared with our proposed method (1.22 $\pm$ 0.36 minutes).

\begin{figure}
\centering
\includegraphics[width=1.0\linewidth]{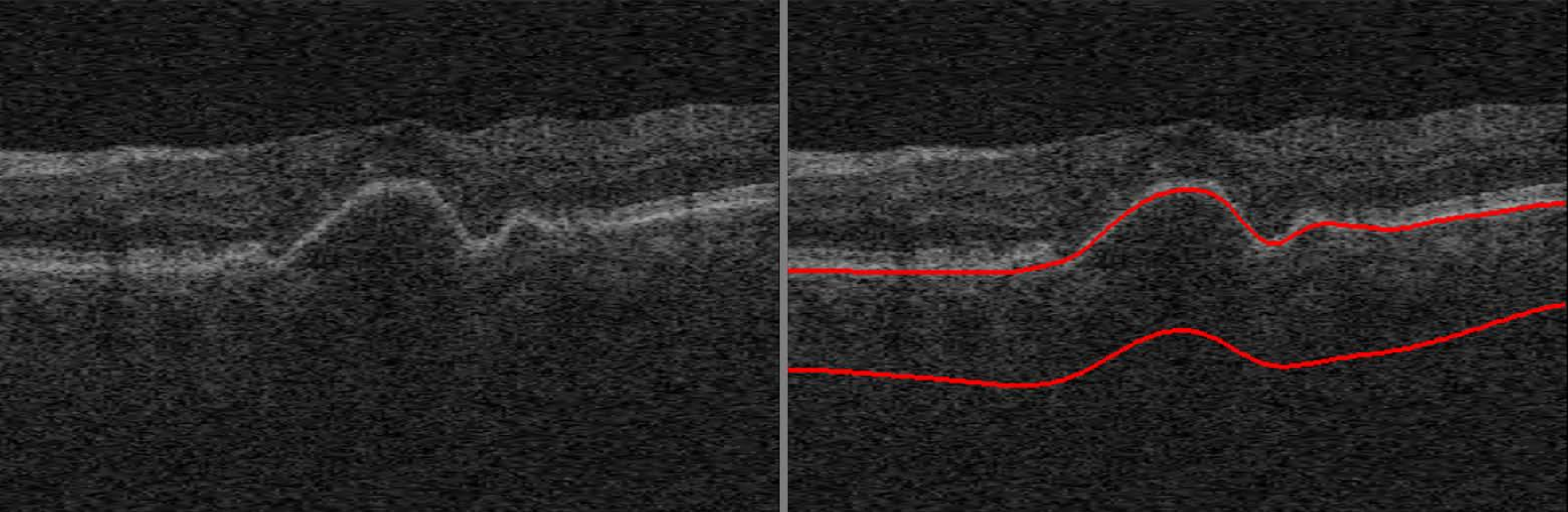}
\caption{Mis-segmented choroidal layer of our proposed method because of pigment epithelial detachment.}
\label{fig.6}
\end{figure}

There are several limitations of the proposed method for automated segmentation of choroidal layers from macular SD-OCT scans. The first is that a relatively small number of OCT scans were used for this study. More OCT scans with the manual tracings of more than 3 B-scans per volume scan from multiple graders are required for more reliable validation of our proposed method. The second one is that the proposed method does not work well for the OCT scans with weak signal in the choroid for which the CSI is not clearly visible by retinal specialists. A study on the association between the SD-OCT signal strength in the choroid and the accuracy of our proposed method would be beneficial. The third is that the proposed method does not work well for OCT scans with large retinal and/or choroidal alterations such as pigment epithelial detachment (Fig.~\ref{fig.6}), choroidal neovascularization, or large drusen. In addition to edge-based and vesselness-based cost functions for our proposed method, deep learning-based cost functions would be helpful for segmentation of the complex choroidal layers, and future work will compare AI-based algorithms to the one proposed here. 

We have proposed a new, automated choroidal layer segmentation method of 3D macular SD-OCT scans and validated it by measuring various error metrics and comparing with the inter-grader variability of two independent, manual graders. For this study, the proposed method showed better performance than our previous method and the inter-grader variability. The proposed method provides a rapid, convenient, automated way of segmenting the choroid from SD-OCT scans, which may be clinically useful for analysis and monitoring of various choroidal diseases.

\section{Acknowledgments}

We appreciate the excellent assistance of Teresa Kopel and Brice Critser in image acquisition and management. This study was supported by the National Institutes of Health grants R01-EY026547, P30-EY025580, R01-EB004640; Research to Prevent Blindness; MDA is the Watzke Professor of Ophthalmology.

\bibliographystyle{unsrt}
\bibliography{References}

\end{document}